\documentclass[preprint,12pt]{elsarticle}

\usepackage{amsmath,amssymb}
\usepackage{graphicx}
\usepackage{booktabs}
\usepackage{hyperref}
\usepackage{xcolor}
\usepackage{multirow}
\usepackage{algorithm}
\usepackage{algorithmic}
\usepackage{url}
\usepackage{subcaption}
\usepackage{tabularx}

\newcommand{\system}{the proposed detector}
\newcommand{\Csystem}{The proposed detector}

\journal{Expert Systems with Applications}

\begin{document}

\begin{frontmatter}

\title{Content-Aware Attack Detection in LLM Agent Tool-Call Traffic: An Empirical Study of Features, Architectures, and Evaluation Protocols}

\author[aff1]{Sultan Zavrak\corref{cor1}}
\ead{sultanzavrak@duzce.edu.tr}
\cortext[cor1]{Corresponding author}
\affiliation[aff1]{organization={Department of Computer Engineering, Duzce University},
  city={Duzce},
  country={Turkey}}

\begin{abstract}
The Model Context Protocol (MCP) has become a widely adopted interface for LLM agents to invoke external tools, yet learned monitoring of MCP tool-call traffic remains underexplored.
In this article, \system{} is presented as an attack detection framework for MCP tool-call traffic that encodes each agent session as a graph (tool calls as nodes, sequential and data-flow links as edges), enriches nodes with sentence-embedding features over arguments and responses, and classifies sessions as benign or attacked.
Three GNN architectures (GAT, GCN, GraphSAGE), a no-graph MLP, and classical baselines (XGBoost, random forest, logistic regression, linear SVM) are evaluated, with the full architecture comparison conducted on RAS-Eval (task-stratified splits) and GraphSAGE retained as the GNN baseline on ATBench and a combined-source variant (both label-stratified).
Three findings emerge.
First, content-level features are essential: metadata-only detection plateaus around an AUROC of 0.64 regardless of architecture, while content embeddings push the AUROC above 0.89.
Second, naive random-split evaluation inflates AUROC by up to 26 percentage points relative to task-disjoint splits, a memorization confound that prior agent-detection work has not addressed.
Third, the detection signal resides primarily in the SBERT content embeddings: an AUROC of 0.975 was reached by tree ensembles on pooled embeddings, performing, for the most part, better than the neural architectures in the primary RAS-Eval setting including GNNs (0.917) and the MLP (0.896), and self-supervised pre-training does not deliver a label-efficiency advantage on this task.
\end{abstract}

\begin{keyword}
attack detection \sep LLM agents \sep Model Context Protocol \sep graph neural networks \sep tool-call security \sep agentic AI safety
\end{keyword}

\end{frontmatter}

%% ============================================================
\section{Introduction}
\label{sec:intro}
%% ============================================================

Large language model (LLM) agents increasingly interact with external tools through standardized protocols.
The Model Context Protocol (MCP) has emerged as a widely adopted interface for this purpose, with over 14{,}000 public repositories tagged \texttt{mcp-server} on GitHub~\cite{github_mcp_topic}, providing agents with access to filesystems, databases, web services, and administrative functions.
Each agent session produces a sequence of tool calls that constitutes an observable attack surface, yet this surface remains largely unmonitored by learned detection systems.

Recent studies show that this attack surface is already practical, not merely hypothetical.
Wang et al.~\cite{wang2025mcptox} demonstrated tool poisoning across 45 MCP servers and 353 tools.
Xiao et al.~\cite{xiao2026mcpsafety} catalogued 20 distinct attack types spanning server, host, and user layers.
Kim et al.~\cite{kim2026sok_agentic} systematized the broader agentic AI threat landscape, identifying tool-mediated attacks as a primary vector.
Despite this growing threat surface, to the author's knowledge, prior learned monitoring of MCP tool-call traffic has not combined externally observable tool-call content and metadata, task-disjoint evaluation, and deployment without access to model internals.

The problem is structurally analogous to network intrusion detection, where flow-based systems abstract raw packet streams into metadata records and flag deviations from learned baselines~\cite{apruzzese2026sok_nids}.
However, the analogy breaks down in practice because MCP attacks primarily alter the \emph{semantic content} of tool interactions while leaving structural metadata unchanged, making metadata-only approaches ineffective for agent traffic.

Recent work has introduced agent behavioral anomaly detectors based on provenance graphs, trajectory verification, and LLM-based oversight~\cite{liu2025traceaegis, trajad2026, mindguard2025, sentinelagent2025}. These approaches, however, require scenario-specific labels, access to model internals, or LLM inference at detection time, as discussed in \S\ref{sec:related}.

In this study, \system{} is presented as a content-aware attack detection framework for MCP tool-call traffic.
\Csystem{} encodes each agent session as a graph where nodes represent individual tool calls and edges capture both sequential ordering and data-flow dependencies.
This representation is evaluated across graph neural network (GNN) architectures, a no-graph multilayer perceptron (MLP), and classical machine learning baselines to isolate what drives detection performance.

The main contributions of this study are summarized as follows:

\begin{enumerate}
    \item \textbf{Session-graph encoding and detector for MCP tool-call traffic.}
    A per-session graph with two edge types (sequential adjacency and data-flow links) and two feature modes (metadata-only and Sentence-BERT (SBERT) content embeddings) is defined and instantiated as a supervised GNN classifier and a self-supervised contrastive variant.

    \item \textbf{Task-stratified evaluation protocol.}
    Standard random-split evaluation is shown to conflate task memorization with attack detection, with the area under the ROC curve (AUROC) inflated by up to 25.8 percentage points under random splits relative to task-disjoint splits. Task-disjoint 70/10/20 splits are adopted as the appropriate protocol for agent attack-detection benchmarks.

    \item \textbf{Empirical findings with deployment implications.}
    Across three GNN architectures, an MLP baseline, four classical classifiers, and three dataset configurations, the experimental results show that content embeddings are necessary, that tree ensembles on pooled SBERT features perform, to a great extent, better than the neural architectures in the primary RAS-Eval setting, and that contrastive pre-training does not provide a reliable label-efficiency advantage.
\end{enumerate}

%% ============================================================
\section{Related Work}
\label{sec:related}
%% ============================================================

The related work is organized along three axes: the agent security threat landscape (\S\ref{sec:rel_threats}), agent-level anomaly detection (\S\ref{sec:rel_agent}), and graph-based attack detection in adjacent domains (\S\ref{sec:rel_graph}).
Table~\ref{tab:related} provides a structured comparison.

\begin{table*}[!htb]
\centering
\caption{Comparison of agent anomaly detection approaches. ``LLM @ inf.'': requires an LLM call at detection time.}
\label{tab:related}
\setlength{\tabcolsep}{4pt}
\renewcommand{\arraystretch}{1.05}
\resizebox{\textwidth}{!}{%
\begin{tabular}{@{}llllccc@{}}
\toprule
\textbf{System} & \textbf{Paradigm} & \textbf{Representation} & \textbf{Model} & \textbf{Labels} & \textbf{MCP} & \textbf{LLM @ inf.} \\
\midrule
TraceAegis~\cite{liu2025traceaegis}        & Supervised & Provenance graph      & Hierarchical templates                            & \checkmark & --         & --         \\
TrajAD~\cite{trajad2026}                   & Supervised & Action trajectory     & Fine-tuned LLM                                    & \checkmark & --         & \checkmark \\
SentinelAgent~\cite{sentinelagent2025}     & LLM-based  & Execution graph       & LLM oversight agent                               & --         & --         & \checkmark \\
Traj.\ Guard~\cite{trajectoryguard2026}    & Rule-based & Action sequence       & Lightweight rules                                 & --         & --         & --         \\
MindGuard~\cite{mindguard2025}             & Supervised & Decision-dep.\ graph  & Attention-based\textsuperscript{\dag}         & \checkmark & \checkmark & --         \\
Temp.\ Attacks~\cite{temporalattack2026}   & Supervised & OpenTelemetry traces           & XGBoost / RF                                      & \checkmark & --         & --         \\
Silent Failures~\cite{silentfailures2025}  & Supervised & OpenTelemetry traces           & XGBoost                                           & \checkmark & --         & --         \\
\midrule
\textbf{This work}                          & \textbf{Sup.\,/\,SSL} & \textbf{Session graph} & \textbf{GAT\,/\,GCN\,/\,SAGE\,/\,MLP}        & \textbf{Varies} & \checkmark & --     \\
\bottomrule
\end{tabular}}%
\\[2pt]
{\footnotesize \textsuperscript{\dag}MindGuard requires access to the agent LLM's attention weights at detection time.}
\end{table*}

\subsection{Agent Security Threat Landscape}
\label{sec:rel_threats}

Kim et al.~\cite{kim2026sok_agentic} provide the most comprehensive systematization of agentic AI threats, categorizing attacks into prompt injection, tool poisoning, data exfiltration, and privilege escalation.
Two benchmarks operationalize this threat model for MCP specifically.
MCPTox~\cite{wang2025mcptox} covers 45~servers, 353~tools, and 1{,}348~test cases across 11~attack categories.
MCP-SafetyBench~\cite{xiao2026mcpsafety} defines 20~attack types across server, host, and user layers with 246~tasks.
Both provide attack taxonomies and test cases but no temporal execution traces suitable for training detection models.

\subsection{Agent Anomaly Detection}
\label{sec:rel_agent}

TraceAegis~\cite{liu2025traceaegis} constructs hierarchical provenance graphs from agent execution traces and validates new traces against learned templates, reporting F1~$>$~0.94 but requiring labeled per-scenario templates.
TrajAD~\cite{trajad2026} trains a specialized LLM to verify trajectories with fine-grained process supervision, which is effective but requires both labeled data and LLM inference at detection time.

MindGuard~\cite{mindguard2025} is the most closely related MCP-specific approach.
It constructs decision-dependence graphs from LLM attention patterns and detects tool poisoning with 94--99\% average precision.
However, it requires access to LLM internals (attention weights), limiting deployment to settings where the agent's underlying model is observable.
\Csystem{} operates on externally observable tool-call metadata and content, requiring no model internals.

SentinelAgent~\cite{sentinelagent2025} deploys an LLM-powered oversight agent over execution graphs for multi-agent systems, introducing an LLM-in-the-loop dependency.
Trajectory Guard~\cite{trajectoryguard2026} provides a lightweight rule-based alternative but cannot generalize beyond predefined patterns.

Two recent works use OpenTelemetry traces for detection.
The Temporal Attack Patterns work~\cite{temporalattack2026} achieves supervised classification on 35k synthetic traces but reports a 66.7\% false positive rate (FPR) on benign data.
Silent Failures detection~\cite{silentfailures2025} applies XGBoost to 16 trace-level features, achieving up to 98\% detection.
Both are supervised and not MCP-specific.

\subsection{Graph-Based Attack Detection in Adjacent Domains}
\label{sec:rel_graph}

Graph neural networks have been applied to attack detection in two adjacent domains.
In network security, E-GraphSAGE~\cite{egraphsage2021} applies GraphSAGE to flow-based network traffic, achieving competitive detection rates while preserving structural information.
In systems security, PROGRAPHER~\cite{prographer2023} uses GNN-based provenance graph embeddings for host-level attack detection, demonstrating that execution-graph structure carries discriminative signal.

This study extends GNN-based classification to a new domain, namely agent-level tool-call graphs, operating at the application layer with semantic content features rather than at the packet level (E-GraphSAGE) or OS level (PROGRAPHER).

\textbf{Position of this study.}
\Csystem{} differs from all of the above in that it operates on externally observable tool-call content (no model internals), supports both supervised and self-supervised training, and is evaluated on real MCP agent traffic with task-disjoint splits.

%% ============================================================
\section{Threat Model}
\label{sec:threat}
%% ============================================================

An LLM agent that interacts with one or more MCP servers during a session is considered, where the agent generates a sequence of tool calls each producing a server response.
\Csystem{} operates as a passive monitor that observes the complete tool-call stream (including tool names, arguments, and responses) without modifying agent behavior.

\subsection{Attack Model}

Attacks in which an adversary manipulates the agent's tool-call behavior through one or more of the following vectors, drawn from the MCP threat taxonomy~\cite{wang2025mcptox, xiao2026mcpsafety, kim2026sok_agentic}, are considered:

\begin{itemize}
    \item \textbf{Tool input manipulation.} The adversary substitutes or modifies tool-call arguments (e.g., replacing a legitimate file path or API identifier with an adversary-supplied one), causing the agent to operate on unintended targets.

    \item \textbf{Tool output manipulation.} A compromised MCP server returns falsified responses (e.g., replacing \texttt{true} with \texttt{false} for a safety check), causing the agent to make incorrect downstream decisions.

    \item \textbf{Combined manipulation.} Both inputs and outputs are manipulated simultaneously, the most common attack mode in practice (81.9\% of attacks in the primary dataset).
\end{itemize}

\subsection{Assumptions}

Access to a corpus of benign agent sessions for training is assumed.
The detector observes full tool-call content (tool names, arguments, and responses) but not the agent's internal reasoning or attention patterns.
This positions \system{} between metadata-only detection (which is shown to be insufficient in \S\ref{sec:paradigm}) and model-internals approaches such as MindGuard~\cite{mindguard2025}, which requires LLM attention weights.

\subsection{Detection Timing}

\Csystem{} operates as a \emph{post-session} auditing detector: it classifies a session after all tool calls have completed.
This reflects the graph-level classification formulation (the full session graph must be constructed before inference) and is appropriate for compliance auditing, incident investigation, and forensic analysis rather than real-time blocking.
Online, per-call detection on partial session graphs is a natural extension but would require incremental graph construction and a decision policy for partial evidence.

%% ============================================================
\section{Detection Framework}
\label{sec:method}
%% ============================================================

\begin{figure*}[!t]
\centering
\includegraphics[width=\textwidth]{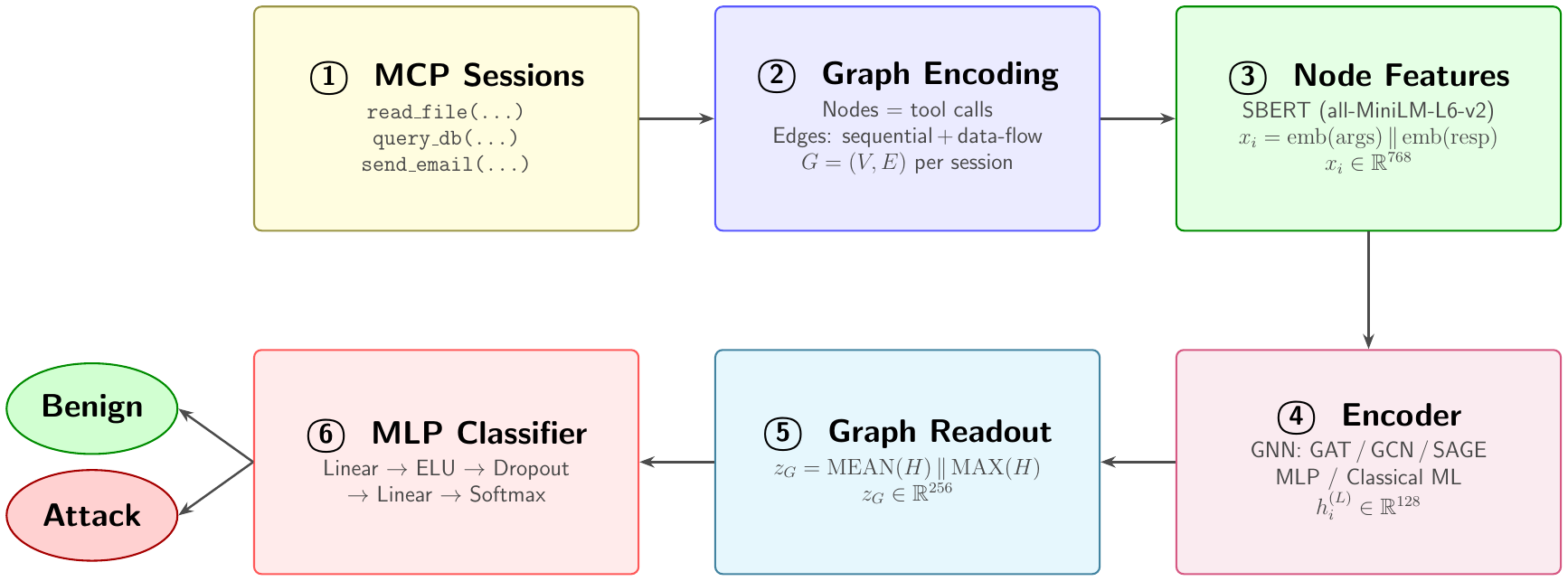}
\caption{Overview of the detection framework. MCP tool-call sessions are encoded as graphs with sequential and data-flow edges, node features are extracted via sentence embeddings, and a classifier (GNN, MLP, or classical ML on pooled features) produces per-session benign/attack predictions.}
\label{fig:architecture}
\end{figure*}

Figure~\ref{fig:architecture} presents an overview of the detection pipeline.

\subsection{Session-to-Graph Encoding}
\label{sec:encoding}

\begin{figure}[!t]
\centering
\includegraphics[width=\columnwidth]{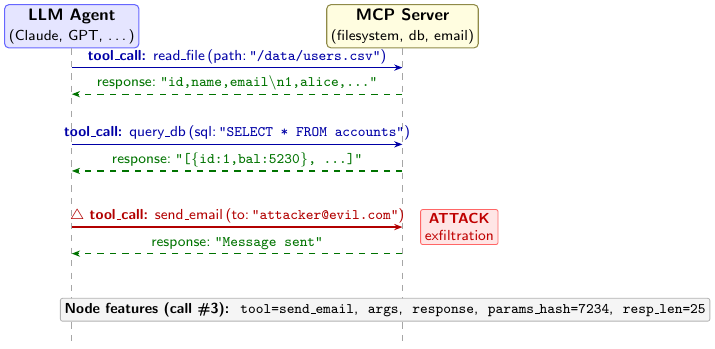}
\caption{Example MCP tool-call session showing a data exfiltration attack: the agent reads sensitive data via \texttt{read\_file} and exfiltrates it via \texttt{send\_email}.}
\label{fig:mcp_traffic}
\end{figure}

\begin{figure}[!t]
\centering
\includegraphics[width=\columnwidth]{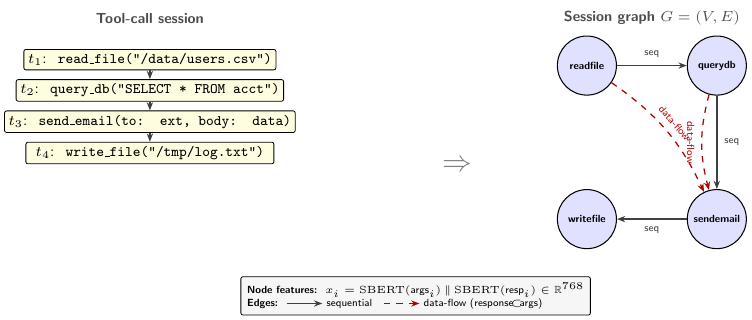}
\caption{Session-to-graph encoding. Left: sequential tool calls. Right: the resulting session graph with sequential edges (solid) and data-flow edges (dashed). Node features are SBERT embeddings of arguments and responses.}
\label{fig:graph_construction}
\end{figure}

Each agent session is encoded as a graph $G = (V, E)$ where nodes represent individual tool calls and edges capture inter-call relationships.
Figure~\ref{fig:mcp_traffic} illustrates a representative MCP session containing a data exfiltration attack, and Figure~\ref{fig:graph_construction} shows the corresponding graph encoding.

\textbf{Nodes.}
For a session with $n$ tool calls, $n$ nodes $v_1, \ldots, v_n$ are created, one per call.
Each node stores the tool name, serialized arguments, tool response, response length, and a parameter hash (MD5 of serialized arguments, modulo 10{,}000).

\textbf{Edges.}
Two edge types are defined, both stored bidirectionally (effectively undirected) to allow symmetric message passing:
\begin{enumerate}
    \item \textbf{Sequential edges:} $v_i \leftrightarrow v_{i+1}$ for consecutive calls, capturing temporal adjacency.
    \item \textbf{Data-flow edges:} $v_i \leftrightarrow v_j$ ($j > i$) when the first 50~characters of the response of call $i$ appear as a substring in the arguments of call $j$, \emph{or} when any whitespace-delimited token of the response (among the first five, length $> 4$) appears in the arguments, capturing explicit data dependencies between tool invocations.
\end{enumerate}
The token-length threshold of 4 suppresses spurious matches on common short tokens while still catching identifier propagation, file paths, and quoted strings. Responses longer than 1{,}000 characters are skipped to avoid false matches on verbose outputs. For single-node graphs (sessions with one tool call), a self-loop is added to ensure a well-defined adjacency structure.
An ablation varying the substring window from 50 to 500 characters produced identical edge counts and AUROC, confirming that the token-level branch subsumes the substring branch on this data.

\subsection{Node Feature Extraction}
\label{sec:features}

Two feature modes are defined to study the importance of semantic content versus structural metadata.

\textbf{Metadata features} ($d = n_\text{tools} + 2$).
Each node receives a one-hot encoding of the tool name ($n_\text{tools}$-dimensional), the normalized parameter hash ($\in [0, 1]$), and the normalized response length (capped at 10{,}000 characters, normalized to $[0, 1]$).

\textbf{Content features} ($d = 768$).
Each node receives the concatenation of two 384-dimensional sentence embeddings: one for the serialized arguments and one for the tool response, both computed using all-MiniLM-L6-v2~\cite{reimers2019sentencebert}.
Input text is truncated to 512~characters.

\textbf{Combined features} ($d = n_\text{tools} + 770$) concatenate both modes.

\subsection{Detection Models}
\label{sec:models}

In order to determine what level of supervision is necessary for MCP attack detection, two detection paradigms are evaluated.

\subsubsection{Supervised GNN Classification}

A two-layer GNN encoder processes node features through message passing:
\begin{equation}
    \mathbf{h}_i^{(l+1)} = \sigma\left(\text{GNN}^{(l)}\left(\mathbf{h}_i^{(l)}, \{\mathbf{h}_j^{(l)} : j \in \mathcal{N}(i)\}\right)\right)
\end{equation}
where $\sigma$ is the ELU activation and $\mathcal{N}(i)$ denotes the neighbors of node $i$.
Three GNN variants are evaluated (Graph Attention Network (GAT)~\cite{velickovic2018gat} with 4 attention heads, Graph Convolutional Network (GCN), and GraphSAGE) to assess architecture sensitivity.
All models are implemented using PyTorch Geometric~\cite{fey2019pyg}.

Graph-level representations are obtained via dual readout:
\begin{equation}
    \mathbf{z}_G = \text{MEAN}(\{\mathbf{h}_i^{(L)}\}) \,\|\, \text{MAX}(\{\mathbf{h}_i^{(L)}\})
\end{equation}
A two-layer MLP classifier maps $\mathbf{z}_G$ to binary output (benign/attacked), trained with cross-entropy loss and inverse-frequency class weights.

\subsubsection{Self-Supervised Learning (SSL): Pre-training + Fine-tuning}

Phase~1 (pre-training on benign data only): a contrastive learning framework generates two augmented views of each graph via node feature masking (rate~0.2) and edge dropping (rate~0.2), and optimizes the Normalized Temperature-scaled Cross-Entropy (NT-Xent) loss with temperature $\tau = 0.5$.
Phase~2 (fine-tuning with labels): the pre-trained encoder is unfrozen and a classification head is trained at a reduced learning rate ($10^{-4}$).
This tests whether benign-only pre-training provides useful initialization.

\subsubsection{MLP Baseline (No Graph Structure)}

In order to isolate the contribution of graph message passing, an MLP baseline is included that applies the same dual readout (mean + max pooling) directly to raw node features, bypassing all GNN layers, followed by the same MLP classifier.

%% ============================================================
\section{Evaluation Protocol}
\label{sec:protocol}
%% ============================================================

\subsection{Datasets}
\label{sec:datasets}

In the experiments carried out in this study, two real-world agent trajectory datasets and a combined-source variant were used, as summarized in Table~\ref{tab:datasets}.

\begin{table}[!htb]
\centering
\caption{Dataset statistics. All sessions contain at least one tool call. The ``Tasks'' column reports the number of distinct task definitions for RAS-Eval (used for task-stratified splits) and the number of curated trajectories for ATBench (which has no shared task structure).}
\label{tab:datasets}
\small
\begin{tabular}{lrrrr}
\toprule
\textbf{Dataset} & \textbf{Benign} & \textbf{Attacked} & \textbf{Tools} & \textbf{Tasks} \\
\midrule
RAS-Eval~\cite{raseval2025} & 605 & 3{,}797 & 34 & 80 \\
ATBench~\cite{atbench2026} & 502 & 497 & 2{,}084 & 999 \\
Combined & 1{,}939 & 4{,}294 & 2{,}118 & --- \\
\bottomrule
\end{tabular}
\end{table}

\textbf{RAS-Eval}~\cite{raseval2025} is the primary dataset used in this study. It provides 80 tasks across 5 domains (calendar, alarm, file management, database, web search) executed by 8 LLM models in a benign setting, yielding 605 benign sessions with at least one tool call (the remaining task--model combinations either produced no tool call or were absent from the released manifest). On the attack side, RAS-Eval supplies 3{,}802 attack tasks (3{,}797 with tool calls) targeting glm-4-flash via three attack modes: tool input manipulation (7.2\%), tool output manipulation (11.1\%), and combined manipulation (81.9\%).
This dataset natively follows the MCP tool-call format, making it the primary MCP-native evaluation target in this study.
The \texttt{guard\_response.jsonl} file (31k guard-model outputs, not agent traces) was excluded.

\textbf{ATBench}~\cite{atbench2026} provides 1{,}000 manually curated trajectories (503~safe, 497~unsafe) across 2{,}084~tools, with attack labels covering multiple risk sources (jailbreak, indirect prompt injection, malicious tool execution, and others). It is important to mention that the ATBench category labels do not align directly with the three vectors defined in the threat model (\S\ref{sec:threat}): jailbreak and indirect prompt injection ultimately surface as input-side manipulation, whereas malicious tool execution typically combines input and output manipulation. ATBench labels are therefore retained as-is for the binary benign/attack classification task, without re-mapping to the three-vector taxonomy. One safe trajectory contains no extractable tool calls and was excluded during graph construction, yielding 502 benign and 497 attack sessions. Sessions are shorter than those in RAS-Eval (median 2--3 tool calls).

\textbf{Combined} merges RAS-Eval benign, ATBench benign, mcpbench benign (832~sessions from cx-cmu/agent\_trajectories), and all attack sessions from RAS-Eval and ATBench, testing multi-source generalization. It should be noted that this configuration is a multi-source variant rather than an independent third benchmark, since the attack sessions are drawn entirely from the same two source datasets and only the benign side is augmented with mcpbench.

\subsection{Task-Stratified Evaluation}
\label{sec:taskstrat}

Standard random-split evaluation allows training and test sets to share the same task definitions.
Because different tasks use different tool subsets, a classifier can learn task-specific tool signatures rather than generalizable attack patterns, which constitutes a form of task memorization.

This confound was discovered empirically: with metadata-only features on RAS-Eval, a random split inflated AUROC by 25.8 percentage points over a task-disjoint split (Table~\ref{tab:leakage}).

\begin{table}[!htb]
\centering
\caption{Task-disjoint vs.\ label-stratified random split on RAS-Eval (supervised GAT, 70/10/20, 3 seeds). The gap between the two protocols quantifies task-memorization inflation. Both protocols use the same model and training pipeline; only the split assignment differs. Content-feature numbers are from the architecture comparison (Table~\ref{tab:arch}) and the matching label-stratified run.}
\label{tab:leakage}
\small
\begin{tabular}{lccc}
\toprule
\textbf{Features} & \textbf{Task-Strat.} & \textbf{Random} & \textbf{Gap} \\
\midrule
Metadata & 0.640 & 0.898 & 0.258 \\
Content & 0.891 & 0.992 & 0.101 \\
\bottomrule
\end{tabular}
\end{table}

Consequently, task-disjoint 70/10/20 splits are adopted as the primary evaluation protocol for RAS-Eval: 70\% of unique tasks are assigned to training, 10\% to validation (early stopping on validation AUROC), and 20\% to testing, with no task overlap across sets.
For ATBench and Combined, which lack a shared task structure, label-stratified 70/10/20 splits are used.

All results are reported as mean~$\pm$~standard deviation across 3~seeds (7, 42, 123).

\subsection{Training Details}

Table~\ref{tab:hyperparams} summarizes the hyperparameters.
All neural models use early stopping on validation AUROC and inverse-frequency class weights.

\begin{table}[!htb]
\centering
\caption{Hyperparameter settings for all experiments.}
\label{tab:hyperparams}
\small
\begin{tabular}{ll}
\toprule
\textbf{Hyperparameter} & \textbf{Value} \\
\midrule
\multicolumn{2}{l}{\textit{Shared}} \\
Hidden dimension & 128 \\
Batch size & 64 \\
Optimizer & Adam \\
Learning rate & $10^{-3}$ \\
Weight decay & $10^{-4}$ \\
Gradient clipping & 1.0 \\
Early stopping patience & 5 (every 10 epochs) \\
Dropout rate & 0.3 \\
Seeds & 7, 42, 123 \\
\midrule
\multicolumn{2}{l}{\textit{Supervised}} \\
Max epochs & 200 \\
GAT attention heads & 4 \\
GNN layers & 2 \\
Readout & mean $\|$ max pooling \\
\midrule
\multicolumn{2}{l}{\textit{Self-supervised pre-training}} \\
Pre-training epochs & 100 \\
Fine-tuning epochs & 50 \\
Fine-tuning learning rate & $10^{-4}$ \\
NT-Xent temperature $\tau$ & 0.5 \\
Node feature mask rate & 0.2 \\
Edge drop rate & 0.2 \\
\midrule
\multicolumn{2}{l}{\textit{Feature extraction}} \\
Sentence encoder & all-MiniLM-L6-v2 \\
Content embedding dim & $2 \times 384 = 768$ \\
Input text truncation & 512 characters \\
Param hash modulus & 10{,}000 \\
Response length cap & 10{,}000 characters \\
\bottomrule
\end{tabular}
\end{table}

\subsection{Metrics}

The following metrics are reported: AUROC as the primary threshold-independent metric, area under the precision-recall curve (AUPRC), macro F1-score, precision, recall, and FPR at threshold 0.5 for supervised models.
Per-attack-mode breakdowns (for RAS-Eval) and per-attack-category breakdowns (for the Combined dataset) report recall and AUROC where the test fold contains both classes.

%% ============================================================
\section{Results}
\label{sec:results}
%% ============================================================

\subsection{Paradigm and Feature Mode Comparison}
\label{sec:paradigm}

Table~\ref{tab:paradigm} presents a comparison of supervised and self-supervised detection with content and metadata features on RAS-Eval under task-stratified evaluation. It is observed that the choice of feature mode has a considerably larger effect on detection performance than the choice of training paradigm.

\begin{table}[!htb]
\centering
\caption{Paradigm and feature-mode comparison on RAS-Eval (task-disjoint splits). Supervised uses GraphSAGE under the 70/10/20 protocol with 3 seeds; SSL+FT uses a GAT encoder for contrastive pre-training followed by fine-tuning, with the AUROC drawn from the full-label point of the label-efficiency study (5-fold cross-validation, Table~\ref{tab:labeleff}). Content features use 768-dim sentence embeddings; metadata uses one-hot tool $+$ params hash $+$ response length. The dashes in the SSL+FT (Content) row reflect that F1, recall, and FPR were not collected at the label-efficiency full-label point under matched threshold settings; the architecture-matched supervised counterpart at full labels is reported in Table~\ref{tab:labeleff}. Architecture-matched SSL vs.\ supervised comparisons should be read against Table~\ref{tab:labeleff} rather than against this table.}
\label{tab:paradigm}
\small
\begin{tabular}{llcccc}
\toprule
\textbf{Method} & \textbf{Features} & \textbf{AUROC} & \textbf{F1} & \textbf{Recall} & \textbf{FPR} \\
\midrule
Sup.\ SAGE     & Content & 0.917{\tiny$\pm$.018} & 0.731{\tiny$\pm$.085} & 0.769{\tiny$\pm$.038} & 0.107{\tiny$\pm$.021} \\
SSL+FT (GAT)   & Content & 0.939{\tiny$\pm$.012} & --- & --- & --- \\
\midrule
Sup.\ SAGE     & Metadata & 0.640{\tiny$\pm$.106} & 0.527{\tiny$\pm$.274} & 0.414{\tiny$\pm$.239} & 0.226{\tiny$\pm$.134} \\
SSL+FT (GAT)   & Metadata & 0.510{\tiny$\pm$.079} & 0.352{\tiny$\pm$.138} & 0.236{\tiny$\pm$.117} & 0.305{\tiny$\pm$.186} \\
\bottomrule
\end{tabular}
\end{table}

Three findings emerge:

\textbf{Finding 1: Content features are essential.}
It can be seen from Table~\ref{tab:paradigm} that content embeddings perform considerably better than metadata-only features across both paradigms.
More specifically, the AUROC of supervised GraphSAGE was lifted from 0.640 (metadata) to 0.917 (content), a 27.7-percentage-point gap.
MCP attacks in RAS-Eval primarily alter the semantic content of tool interactions (substituting arguments, returning falsified responses) rather than changing which tools are called or how they are structured.
A metadata-only detector cannot recover this signal regardless of model sophistication.

\textbf{Finding 2: SSL matches but does not surpass supervised under matched architecture, and only at the full label budget.}
Under architecture-matched comparison (GAT encoder, 5-fold cross-validation), an AUROC of 0.939 was reached by contrastive pre-training followed by fine-tuning, which is statistically indistinguishable from supervised training from scratch with the same encoder (0.944, Table~\ref{tab:labeleff}). When compared with the strongest supervised configuration (GraphSAGE, 0.917 in Table~\ref{tab:paradigm}), SSL+FT (GAT) lies in the same range, although the comparison is then confounded by architecture choice.
This parity, however, does not extend to the low-label regime (\S\ref{sec:labeleff}). It should be noted that supervised training matched or performed better than SSL at every label fraction $\geq 5\%$.
With metadata features, SSL collapsed to an AUROC of 0.510, confirming that content-level signal is what makes the contrastive objective useful.

\textbf{Finding 3: Metadata features enable task memorization.}
As shown in Table~\ref{tab:leakage}, AUROC is inflated by 25.8 percentage points under random splits when metadata features are used (0.898 vs. 0.640), whereas a much smaller 10.1-percentage-point gap is observed for content features (0.992 vs. 0.891).
This suggests that metadata-based classifiers learn task-specific tool signatures from training data, whereas content features capture generalizable semantic differences between benign and attacked interactions.

\subsection{Architecture Comparison}
\label{sec:arch}

Table~\ref{tab:arch} compares the three GNN architectures and the no-graph MLP baseline on RAS-Eval with content features under task-stratified evaluation.

\begin{table}[!htb]
\centering
\caption{Architecture comparison on RAS-Eval (content features, task-stratified, 3 seeds). MLP applies the same readout (mean + max pooling) to raw node features without graph convolutions.}
\label{tab:arch}
\small
\begin{tabular}{lcccc}
\toprule
\textbf{Architecture} & \textbf{AUROC} & \textbf{F1} & \textbf{Recall} & \textbf{FPR} \\
\midrule
GraphSAGE & \textbf{0.917}{\tiny$\pm$.018} & 0.731{\tiny$\pm$.085} & 0.769{\tiny$\pm$.038} & \textbf{0.107}{\tiny$\pm$.021} \\
GCN & 0.902{\tiny$\pm$.007} & 0.740{\tiny$\pm$.054} & 0.789{\tiny$\pm$.028} & 0.131{\tiny$\pm$.019} \\
MLP (no graph) & 0.896{\tiny$\pm$.010} & 0.727{\tiny$\pm$.066} & 0.773{\tiny$\pm$.033} & 0.143{\tiny$\pm$.056} \\
GAT & 0.891{\tiny$\pm$.023} & \textbf{0.767}{\tiny$\pm$.028} & \textbf{0.832}{\tiny$\pm$.052} & 0.178{\tiny$\pm$.046} \\
\bottomrule
\end{tabular}
\end{table}

As can be seen from Table~\ref{tab:arch}, an AUROC above 0.89 was attained by all architectures, with differences of at most 2.6 percentage points across models. It is realized that the benefit of graph message passing over the no-graph MLP is architecture-dependent: GraphSAGE and GCN slightly exceed the MLP, whereas GAT falls marginally below it. The dominant signal is therefore content-level rather than structural.

\subsection{Classical Baselines}
\label{sec:baselines}

In order to further isolate the contribution of model architecture from feature quality, classical machine learning classifiers were evaluated on the same pooled SBERT features used by the MLP baseline (1536-dimensional mean + max pooling of per-node content embeddings).
All classifiers use fixed hyperparameters with class weighting: logistic regression (L-BFGS, $C\!=\!1.0$, balanced weights), linear support vector machine (SVM) via stochastic gradient descent (SGD) (hinge loss, $\alpha\!=\!10^{-4}$, balanced weights), random forest (200~trees, balanced weights), and XGBoost (200~rounds, histogram splitting, \texttt{scale\_pos\_weight} set to the benign/attack ratio). For the linear SVM, the AUROC is computed from the raw decision-function score (the signed distance to the hyperplane) rather than from a calibrated probability, which is the standard convention for hinge-loss classifiers.
No hyperparameter tuning was performed.
Table~\ref{tab:baselines} presents results across all three datasets.

\begin{table}[!htb]
\centering
\caption{Classical baselines on pooled content features (3 seeds). RAS-Eval uses task-stratified splits; ATBench and Combined use label-stratified splits. All classifiers use class weighting for imbalance.}
\label{tab:baselines}
\footnotesize
\begin{tabular}{llcccc}
\toprule
\textbf{Dataset} & \textbf{Classifier} & \textbf{AUROC} & \textbf{F1} & \textbf{Recall} & \textbf{FPR} \\
\midrule
\multirow{4}{*}{RAS-Eval}
& XGBoost & \textbf{0.975}{\tiny$\pm$.005} & 0.958{\tiny$\pm$.005} & 0.927{\tiny$\pm$.008} & \textbf{0.071}{\tiny$\pm$.031} \\
& Random Forest & 0.974{\tiny$\pm$.006} & \textbf{0.972}{\tiny$\pm$.005} & \textbf{0.966}{\tiny$\pm$.015} & 0.191{\tiny$\pm$.052} \\
& Logistic Reg. & 0.917{\tiny$\pm$.022} & 0.920{\tiny$\pm$.014} & 0.865{\tiny$\pm$.023} & 0.128{\tiny$\pm$.050} \\
& Linear SVM & 0.891{\tiny$\pm$.040} & 0.858{\tiny$\pm$.078} & 0.777{\tiny$\pm$.137} & 0.180{\tiny$\pm$.064} \\
\midrule
\multirow{4}{*}{ATBench}
& XGBoost & 0.758{\tiny$\pm$.015} & 0.668{\tiny$\pm$.027} & 0.643{\tiny$\pm$.041} & 0.274{\tiny$\pm$.005} \\
& Random Forest & \textbf{0.784}{\tiny$\pm$.006} & \textbf{0.669}{\tiny$\pm$.036} & 0.613{\tiny$\pm$.055} & \textbf{0.211}{\tiny$\pm$.033} \\
& Logistic Reg. & 0.743{\tiny$\pm$.017} & 0.662{\tiny$\pm$.015} & \textbf{0.653}{\tiny$\pm$.034} & 0.314{\tiny$\pm$.045} \\
& Linear SVM & 0.729{\tiny$\pm$.027} & 0.627{\tiny$\pm$.049} & 0.599{\tiny$\pm$.093} & 0.294{\tiny$\pm$.061} \\
\midrule
\multirow{4}{*}{Combined}
& XGBoost & \textbf{0.986}{\tiny$\pm$.001} & \textbf{0.952}{\tiny$\pm$.001} & \textbf{0.944}{\tiny$\pm$.001} & \textbf{0.086}{\tiny$\pm$.007} \\
& Random Forest & 0.982{\tiny$\pm$.002} & 0.942{\tiny$\pm$.003} & 0.928{\tiny$\pm$.005} & 0.093{\tiny$\pm$.000} \\
& Logistic Reg. & 0.956{\tiny$\pm$.010} & 0.940{\tiny$\pm$.003} & 0.940{\tiny$\pm$.005} & 0.132{\tiny$\pm$.003} \\
& Linear SVM & 0.952{\tiny$\pm$.007} & 0.933{\tiny$\pm$.002} & 0.940{\tiny$\pm$.003} & 0.168{\tiny$\pm$.003} \\
\bottomrule
\end{tabular}
\end{table}

The obtained results indicate that the tree-based ensembles were, for the most part, stronger than the neural architectures, with XGBoost and random forest yielding 6--8 percentage points higher AUROC than the best GNN on RAS-Eval while remaining competitive on ATBench and Combined, where the margins were smaller.
In addition to this, logistic regression matched the best GNN on RAS-Eval (0.917), confirming that the detection signal resides in the pooled SBERT embeddings and does not require graph-aware models.

\subsection{Per-Dataset Results}
\label{sec:perdataset}

Table~\ref{tab:perdataset} presents results for each dataset configuration.

\begin{table}[!htb]
\centering
\caption{Per-dataset results (content features, GraphSAGE, 3 seeds). RAS-Eval uses task-stratified splits; ATBench and Combined use label-stratified splits.}
\label{tab:perdataset}
\small
\begin{tabular}{lcccc}
\toprule
\textbf{Dataset} & \textbf{AUROC} & \textbf{AUPRC} & \textbf{Recall} & \textbf{FPR} \\
\midrule
RAS-Eval & 0.917{\tiny$\pm$.018} & 0.974{\tiny$\pm$.022} & 0.769{\tiny$\pm$.038} & 0.107{\tiny$\pm$.021} \\
ATBench & 0.762{\tiny$\pm$.029} & 0.758{\tiny$\pm$.012} & 0.643{\tiny$\pm$.062} & 0.264{\tiny$\pm$.026} \\
Combined & \textbf{0.971}{\tiny$\pm$.004} & \textbf{0.987}{\tiny$\pm$.001} & \textbf{0.917}{\tiny$\pm$.016} & 0.098{\tiny$\pm$.014} \\
\bottomrule
\end{tabular}
\end{table}

As can be seen from Table~\ref{tab:perdataset}, the three dataset configurations exhibit markedly different difficulty profiles. The strongest task-stratified detection is observed on RAS-Eval, which benefits from a consistent MCP-native format and well-defined attack modes, whereas ATBench proves the hardest, owing to its shorter sessions (median 2--3 tool calls) and semantically subtle attacks that provide less content-level signal. The Combined configuration yields the highest overall performance under label-stratified splits, but because mcpbench contributes only benign examples from a different agent framework, source-specific cues may be partially exploited by the model. This result is therefore interpreted as mixed-source in-distribution performance rather than evidence that pooling benign sources improves detection in general.

\subsection{Per-Attack-Mode Analysis}
\label{sec:perattack}

Table~\ref{tab:perattack} breaks down detection performance by attack mode on RAS-Eval.

\begin{table}[!htb]
\centering
\caption{Per-attack-mode detection on RAS-Eval (supervised GraphSAGE, content features, task-stratified, seed=123). Sample counts $N$ are the test-fold totals for this seed and therefore differ slightly from the global mode distribution reported in \S\ref{sec:datasets}. The single-seed weighted recall implied by this table (approximately 0.82) is higher than the 3-seed mean recall reported in Table~\ref{tab:perdataset} (0.769), reflecting per-seed variance on a relatively small test fold.}
\label{tab:perattack}
\small
\begin{tabular}{lrr}
\toprule
\textbf{Attack Mode} & \textbf{N} & \textbf{Recall} \\
\midrule
tool\_input + tool\_output & 527 & 0.869 \\
tool\_output only & 77 & 0.818 \\
tool\_input only & 57 & 0.368 \\
\bottomrule
\end{tabular}
\end{table}

It can be seen from Table~\ref{tab:perattack} that detection recall scales with the number of manipulated channels. Combined manipulation, which alters both arguments and responses, is the most detectable mode because both content streams deviate from benign patterns, providing multiple content-level signals.
Bearing this in mind, input-only manipulation proved to be the hardest to detect (recall of 0.368). Substituting a single argument value (e.g., a different arXiv ID or file path) produces a small change in the argument embedding that may fall within normal variation.

\begin{figure}[!t]
\centering
\includegraphics[width=\columnwidth]{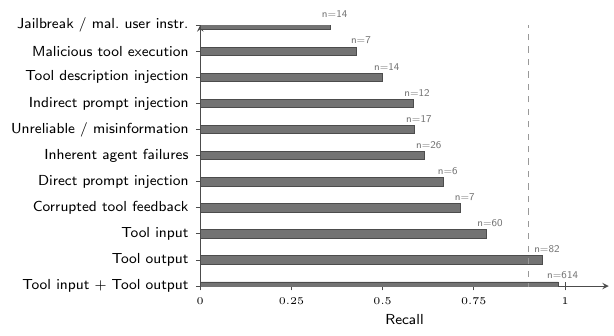}
\caption{Per-attack-category recall on the Combined dataset (GraphSAGE, content features). Categories with both tool input and output manipulation are the most detectable. Semantic-level attacks (jailbreak, misinformation) are the hardest.}
\label{fig:per_attack}
\end{figure}

Figure~\ref{fig:per_attack} provides a finer-grained breakdown across all attack categories on the Combined dataset.

\subsection{Label Efficiency}
\label{sec:labeleff}

In order to assess practical deployment requirements, SSL pre-training was compared against supervised training from scratch at varying label fractions (1\%--100\%), using 5-fold cross-validation on all three datasets (task-disjoint folds for RAS-Eval, label-stratified for ATBench and Combined).
Here, SSL denotes contrastive learning on benign data followed by fine-tuning.
Both paradigms use a GAT encoder to isolate the paradigm effect from architecture choice.

\begin{table}[!htb]
\centering
\caption{Label efficiency: AUROC at varying label fractions (5-fold cross-validation). RAS-Eval uses task-disjoint folds; ATBench and Combined use label-stratified folds.}
\label{tab:labeleff}
\small
\begin{tabular}{lcccccc}
\toprule
& \multicolumn{2}{c}{\textbf{RAS-Eval}} & \multicolumn{2}{c}{\textbf{ATBench}} & \multicolumn{2}{c}{\textbf{Combined}} \\
\cmidrule(lr){2-3}\cmidrule(lr){4-5}\cmidrule(lr){6-7}
\textbf{Labels} & SSL+FT & Sup & SSL+FT & Sup & SSL+FT & Sup \\
\midrule
1\% & .547{\tiny$\pm$.159} & .584{\tiny$\pm$.135} & \textbf{.530}{\tiny$\pm$.049} & .508{\tiny$\pm$.046} & .789{\tiny$\pm$.028} & .798{\tiny$\pm$.027} \\
5\% & .556{\tiny$\pm$.103} & \textbf{.707}{\tiny$\pm$.120} & .553{\tiny$\pm$.039} & \textbf{.615}{\tiny$\pm$.052} & .877{\tiny$\pm$.014} & \textbf{.884}{\tiny$\pm$.016} \\
10\% & .686{\tiny$\pm$.082} & \textbf{.800}{\tiny$\pm$.076} & .545{\tiny$\pm$.053} & \textbf{.635}{\tiny$\pm$.040} & .917{\tiny$\pm$.008} & \textbf{.927}{\tiny$\pm$.016} \\
25\% & .796{\tiny$\pm$.115} & \textbf{.827}{\tiny$\pm$.102} & .631{\tiny$\pm$.043} & \textbf{.681}{\tiny$\pm$.031} & \textbf{.956}{\tiny$\pm$.006} & .956{\tiny$\pm$.004} \\
50\% & .914{\tiny$\pm$.037} & \textbf{.929}{\tiny$\pm$.030} & .708{\tiny$\pm$.036} & \textbf{.728}{\tiny$\pm$.041} & \textbf{.966}{\tiny$\pm$.005} & .965{\tiny$\pm$.006} \\
100\% & .939{\tiny$\pm$.012} & \textbf{.944}{\tiny$\pm$.016} & .730{\tiny$\pm$.023} & \textbf{.733}{\tiny$\pm$.027} & \textbf{.971}{\tiny$\pm$.005} & .971{\tiny$\pm$.006} \\
\bottomrule
\end{tabular}
\end{table}

\begin{figure*}[!t]
\centering
\includegraphics[width=\textwidth]{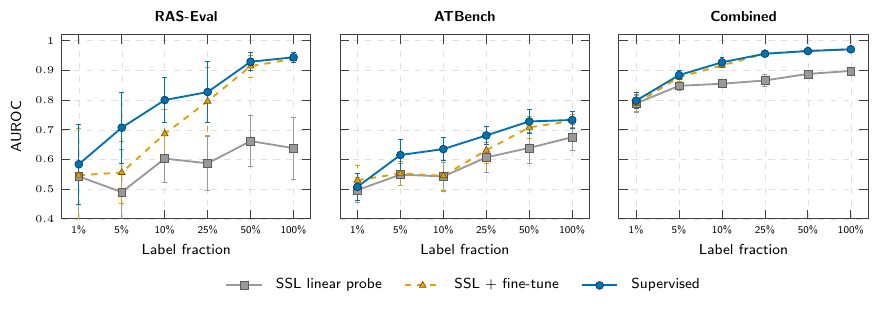}
\caption{Label efficiency across datasets. Supervised training (circles) matches or performs better than SSL fine-tuning (triangles) at all label fractions beyond 1\%, with the two methods converging at the full label budget on every dataset.}
\label{fig:label_efficiency}
\end{figure*}

As shown in Table~\ref{tab:labeleff}, an AUROC of 0.800 was reached by supervised training from scratch with 10\% of labels on RAS-Eval, and 0.927 on Combined.
SSL pre-training provided a marginal advantage only at 1\% labels on ATBench (0.530 vs. 0.508). From 5\% labels upward, supervised consistently matched or performed better than SSL, with the two methods converging at 100\% labels on all datasets.
These results indicate that supervised training is sufficient whenever even modest annotation is available, and that contrastive pre-training on benign sessions does not provide a meaningful encoder initialization for this task.
Figure~\ref{fig:label_efficiency} visualizes these trends.

%% ============================================================
\section{Discussion}
\label{sec:discussion}
%% ============================================================

\subsection{The Role of Graph Structure}

Based on the detailed discussion of Tables~\ref{tab:arch} and~\ref{tab:baselines}, it is observed that the no-graph MLP nearly matched the best GNN, while tree ensembles on the same pooled SBERT features yielded higher performance than all neural architectures in the primary RAS-Eval setting.
The dominant detection signal therefore resides in the content-level sentence embeddings, not in graph structure or model architecture.
The graph representation nonetheless provides a natural framework for incorporating richer structural information (such as tool dependency graphs or temporal dynamics) where the structural signal may be stronger than in the current datasets.

\subsection{Comparison with MindGuard}

An average precision of 94--99\% on MCP tool poisoning was reported by MindGuard~\cite{mindguard2025} using decision-dependence graphs derived from LLM attention weights. Direct numerical comparison is not possible due to different datasets and threat models.
The key difference is that MindGuard requires model internals, while \system{} operates on externally observable tool-call content, making it deployable as a gateway-level monitor. This comes at the cost of lower detection granularity for attacks that do not alter observable behavior.

\subsection{Limitations}

\textbf{Dataset scope.}
Both datasets are research benchmarks, and real-world MCP attacks observed in production may exhibit characteristics absent from either source. This limitation is mitigated by reporting on the combined-source variant in addition to each dataset alone.

\textbf{Content dependency.}
The reliance on content-level features (sentence embeddings of tool arguments and responses) means \system{} requires access to tool-call content, not just metadata.
In privacy-sensitive deployments where content inspection is restricted, the metadata-only mode (AUROC of 0.64) may be the only viable option.

\textbf{Single-model attacks.}
RAS-Eval benign sessions span 8 LLMs but the attack subset targets only glm-4-flash.
This is by design (the released attack set was generated against a single model).
A metadata-level confound check was carried out: a 5-fold benign-only classifier predicting whether a session was generated by glm-4-flash achieved an AUROC of 0.41 (below chance), confirming that metadata features cannot fingerprint the underlying model.
However, this does not rule out content-level leakage: SBERT embeddings of glm-4-flash responses may cluster differently from other models', and the detector could partially exploit this stylistic signal.
A same-model-only evaluation would isolate attack signal from model identity but is not feasible with the current release (too few benign glm-4-flash sessions). Consequently, evaluation on attacks targeting diverse models remains the most direct way to strengthen generalizability.

\textbf{Sentence embedding model.}
All content features use all-MiniLM-L6-v2, a general-purpose sentence encoder.
A domain-adapted encoder trained on tool-call text could improve detection, but this is left to future work to maintain reproducibility with off-the-shelf components.

\subsection{Deployment Considerations}

\Csystem{} is designed for passive deployment as a monitoring layer alongside MCP gateways.
At inference time the dominant cost is sentence embedding computation. This cost is bounded by the per-input inference cost of all-MiniLM-L6-v2~\cite{reimers2019sentencebert} and can be cached per unique tool-call to amortize the cost across repeated invocations.
The label-efficiency results (\S\ref{sec:labeleff}) indicate that labeling as few as 10\% of sessions reaches roughly 85--96\% of the full-budget AUROC, making initial deployment practical with modest annotation effort.

%% ============================================================
\section{Conclusion}
\label{sec:conclusion}
%% ============================================================

In this study, \system{} was presented as a content-aware attack detector for MCP tool-call traffic, and was used to study what carries the detection signal at the session level.
It can be concluded that three findings emerge from the experimental results.
First, the dominant signal is found to reside in the semantic content of arguments and responses, and structural metadata alone is insufficient regardless of architecture.
On RAS-Eval, an AUROC of 0.975 was reached by tree ensembles on pooled SBERT embeddings, an outcome that, by and large, surpassed the neural architectures and confirms that the value of \system{} lies in the content-aware feature representation rather than the classification architecture.
Second, it is important to mention that naive random-split evaluation conflates task memorization with attack detection. A task-stratified protocol is necessary for honest benchmarking of agent-level attack detection.
Third, contrastive pre-training does not, on the whole, deliver the label-efficiency advantage commonly attributed to SSL on this task, and supervised training from scratch is, for the most part, at least as strong below the full label budget.

It remains to be seen whether temporal graph modeling for session-level dynamics will further improve detection. Future work will also evaluate \system{} on real-world MCP attack logs as they become available, and investigate federated deployment for cross-organization monitoring without centralizing sensitive tool-call data.

%% ============================================================
\section*{CRediT authorship contribution statement}
%% ============================================================

\textbf{Sultan Zavrak:} Conceptualization, Methodology, Software, Validation, Formal analysis, Investigation, Writing -- original draft, Writing -- review \& editing, Visualization.

%% ============================================================
\section*{Declaration of competing interest}
%% ============================================================

The author declares that there are no known competing financial interests or personal relationships that could have appeared to influence the work reported in this paper.

%% ============================================================
\section*{Data availability}
%% ============================================================

The datasets used in this study are publicly available.
RAS-Eval~\cite{raseval2025} is available on GitHub as part of the RAS-Eval benchmark.
ATBench~\cite{atbench2026} is available on HuggingFace (AI45Research/ATBench).
The mcpbench subset is available on HuggingFace as part of cx-cmu/agent\_trajectories.

%% ============================================================
\section*{Code availability}
%% ============================================================

The source code for all experiments, including graph construction, model training, evaluation scripts, and configuration files, is available at: \url{https://github.com/szavrak/mcp-agent-attack-detection}.

%% ============================================================
\section*{Declaration of generative AI and AI-assisted technologies in the writing process}
%% ============================================================

During the preparation of this work the author used ChatGPT, Claude, and Gemini to assist with coding, drafting, writing, and polishing of the manuscript. After using these tools, the author reviewed and edited the content as needed and takes full responsibility for the content of the published article.

%% ============================================================
\section*{Acknowledgements}
%% ============================================================

Computational experiments were performed using Google Colab with NVIDIA T4 GPU resources.

%% ============================================================
%% References
%% ============================================================

\bibliographystyle{elsarticle-num}
\bibliography{../references}

\end{document}